\documentstyle[12pt,amssymb]{article}

\def\fnote#1#2{\begingroup\def\thefootnote{#1}\footnote{#2}\addtocounter{footnote}{-1}\endgroup}

\def\inbar{\vrule height1.5ex width.4pt depth0pt}
\def\IB{\relax{\rm I\kern-.18em B}}
\def\IC{\relax\,\hbox{$\inbar\kern-.3em{\rm C}$}}
\def\ID{\relax{\rm I\kern-.18em D}}
\def\IE{\relax{\rm I\kern-.18em E}}
\def\IF{\relax{\rm I\kern-.18em F}}
\def\IG{\relax\,\hbox{$\inbar\kern-.3em{\rm G}$}}
\def\IH{\relax{\rm I\kern-.18em H}}
\def\II{\relax{\rm I\kern-.18em I}}
\def\IK{\relax{\rm I\kern-.18em K}}
\def\IL{\relax{\rm I\kern-.18em L}}
\def\IM{\relax{\rm I\kern-.18em M}}
\def\IN{\relax{\rm I\kern-.18em N}}
\def\IO{\relax\,\hbox{$\inbar\kern-.3em{\rm O}$}}
\def\IP{\relax{\rm I\kern-.18em P}}
\def\IQ{\relax\,\hbox{$\inbar\kern-.3em{\rm Q}$}}
\def\IR{\relax{\rm I\kern-.18em R}}
\def\IT{\relax{\rm I\kern-.18em T}}
\def\ZZ{\relax{\sf Z\kern-.4em Z}}

\def\a{\alpha}       \def\g{\gamma}  
\def\e{\epsilon} \def\G{\Gamma}     
    \def\Om{\Omega}

\def\cO{{\cal O}} \def\cP{{\cal P}}  
  \def\cU{{\cal U}}

 \def\pfrak{{\mathfrak p}}

   \def\mathF{{\mathbb F}}  \def\mathG{{\mathbb G}}
 \def\mathN{{\mathbb N}}  \def\mathP{{\mathbb P}}  \def\mathQ{{\mathbb Q}}
   \def\mathZ{{\mathbb Z}}

\def\fnote#1#2{\begingroup\def\thefootnote{#1}\footnote{#2}\addtocounter
{footnote}{-1}\endgroup}
\def\beq{\begin{equation}}
\def\eeq{\end{equation}}
\def\bea{\begin{eqnarray}}
\def\eea{\end{eqnarray}}
\def\llea#1{\label{#1}\eea}
\def\lleq#1{\label{#1}\eeq}
\let\nn=\nonumber
\def\tabroom{\hbox to0pt{\phantom{\Huge A}\hss}}
\def\notin{\ \hbox{{$\in$}\kern-.51em\hbox{/}}}

\def\lra{\longrightarrow}

  \def\E1Fq{E_1/\IF_q}

\def\uu{{\underline u}}

\def\rmA{{\rm A}} \def\rmB{{\rm B}} \def\rmC{{\rm C}}
  \def\rmK{{\rm K}}
 \def\rmN{{\rm N}}

\def\rmaff{{\rm aff}}

        \def\rmdeg{{\rm deg}}
   \def\rmdim{{\rm dim}}
\def\rmell{{\rm ell}}   
   
\def\rmlcm{{\rm lcm}}
\def\rmmod{{\rm mod}}

\def\rmrk{{\rm rk}}
\def\rmsign{{\rm sign}}
        
\def\rmwt{{\rm wt}}

   \def\rmBP{{\rm BP}}
     
       \def\rmGal{{\rm Gal}}
     \def\rmGr{{\rm Gr}}

\def\rmSL{{\rm SL}}      \def\rmSU{{\rm SU}}
\def\rmTr{{\rm Tr}}

\def\notdiv{{\relax{~|\kern-.35em /~}}}

\def\boxit#1{
\vbox{\hrule height1pt\hbox{\vrule width1pt\kern0.3cm
\vbox{\kern0.3cm\hbox{$\displaystyle#1$}\kern0.3cm}\kern0.3cm\vrule
width1pt}\hrule height1pt}}

\begin{document}
\parindent=0pt

\phantom{\hfill Version 9, \today}

%{TO DO: $5^2,7^2$-terms; conifolds and other types (I/II); degeneration behavior;}

\vskip 0.4truein

 \centerline{{\bf STRING MODULAR PHASES IN CALABI-YAU FAMILIES%\fnote{\dagger}{This paper is dedicated
            % to the memory of Max Kreuzer.}
            }}

\vskip .2truein

 \centerline{\sc Shabnam Kadir\fnote{1}{email: kadir@math.uni-hannover.de},
                 Monika Lynker\fnote{2}{email: mlynker@iusb.edu}
                 and Rolf Schimmrigk\fnote{2}{email: netahu@yahoo.com, rschimmr@iusb.edu}}

\vskip .2truein

 \centerline{ $^1$Institut f\"ur Algebraische Geometrie} \vskip .03truein
 \centerline{ Leibniz Universit\"at Hannover}\vskip .03truein
 \centerline{ Welfengarten 1, 30167 Hannover}

\vskip .2truein

 \centerline{ $^2$Department of Physics} \vskip .03truein
 \centerline{ Indiana University South Bend} \vskip .03truein
\centerline{ 1700 Mishawaka Ave., South Bend, IN 46634}

\vskip .3truein

\centerline{\it Dedicated to the memory of Max Kreuzer}

\vskip .4truein

\baselineskip=16pt

\centerline{\bf Abstract:}

\vskip .03truein

 \begin{quote}
 We investigate the structure of singular Calabi-Yau varieties in moduli spaces that
 contain a Brieskorn-Pham point. Our main tool is a construction of families
 of deformed motives over the parameter space. We analyze these motives for general
 fibers and explicitly compute the $L-$series for singular fibers for several families.
 We find that the resulting motivic $L-$functions agree with the $L-$series of modular
 forms whose weight depends both on the rank of the motive and the degree of the
 degeneration of the variety. Surprisingly, these motivic $L-$functions are identical
 in several cases to $L-$series derived from weighted Fermat hypersurfaces. This shows
 that singular Calabi-Yau spaces of non-conifold type can admit a string worldsheet
 interpretation, much like rational theories, and that the corresponding irrational
 conformal field theories inherit information from the Gepner conformal field theory
 of the weighted Fermat fiber of the family. These results suggest that phase transitions
 via non-conifold configurations are physically plausible. In the case of severe
 degenerations we find a dimensional transmutation of the motives. This suggests
 further that singular configurations with non-conifold singularities may facilitate
 transitions between Calabi-Yau varieties of different dimensions.
 \end{quote}

\vfill

\renewcommand\thepage{}
\newpage

\baselineskip=15pt
 \parskip=1pt

 \tableofcontents

\vskip .5truein

% ar:    22 / st: 20.7  / bh:  22.2 / cm:  20.5 /  gkm: 22 /
% arith: 19 / k3: 17.2  / ems: 22   / ams:      /  es:  18 /

\baselineskip=19.3pt

\parskip=.16truein
\parindent=0pt
\pagenumbering{arabic}

\section{Introduction and outline}

 Previous work has provided support for the idea that the problem of deriving the geometry
 of spacetime in string theory from first principles can be approached via the
 concept of automorphic motives. The strategy of this framework is to relate
 automorphic forms derived from the geometry of spacetime to automorphic forms
 determined by the two-dimensional conformal field theory on the string worldsheet.
 It is possible to invert this process, and to construct the geometry of spacetime from
 the string theoretic automorphic forms. More precisely, it is possible to derive
 motivic pieces of the compactification variety from forms determined by the
 worldsheet theory. The combination of the resulting motivic building
 blocks then determines the global structure of the Calabi-Yau space.
 This approach thus leads to an explicit and computable realization of the concept
 of an emergent spacetime in string theory. The strategy outlined above has been
 pursued in the recent past in the context of weighted Fermat hypersurfaces and
 Gepner models in a number of papers (see e.g. \cite{rs06,rs08,kls08} and references
 therein).

 Weighted Fermat hypersurfaces, or Brieskorn-Pham varieties, are special in
 that the underlying conformal field theory is rational. Rationality of the theory
 implies that the algebra and its representations are under precise control
 and it is possible to understand the relationship between the worldsheet theory
 and the geometry of Calabi-Yau manifolds in some detail. In the early years
 after the original construction of Gepner \cite{g88} the techniques applied to this end
  came from Landau-Ginzburg theories \cite{m88,vw88,lvw89} and non-linear sigma
  models  \cite{w93}.
 The arithmetic geometric approach pursued in \cite{rs06,rs08} in the context of
  weighted Fermat hypersurfaces and Gepner models relates the geometric
  modular forms derived from motives of the Calabi-Yau varieties to Hecke
  indefinite modular forms. These in turn are built from
 the string functions that appear in the partition function of the exactly solvable model. This method
 thereby establishes a close connection between modular forms of rather different origins.

 Since the constructions of Gepner and Kazama-Suzuki two decades
 ago \cite{g88,ks89}, it has proven difficult to extend the detailed understanding
 obtained for rational theories to deformations of such theories.
 Much less, therefore, is known about the conformal field theories that correspond
 to families of Calabi-Yau varieties.
 Our purpose in this paper is to initiate a program to address this problem
 by extending the modularity methods used for K3 surfaces and Calabi-Yau threefolds of
  Brieskorn-Pham type in \cite{rs06,rs08} to the class of Calabi-Yau families which
  contain a Brieskorn-Pham fiber in their moduli space.
  The structure of such families is of the form
 \beq
  X_n^d(\psi_D) = \left\{(z_0:\cdots :z_{n+1}) \in \mathP_{(w_0,...,w_{n+1})}
   ~{\Big |}~ P(z_i,\psi_D) ~=~ p_\rmBP(z_i)  - \sum_{J\in D} \psi_J z^J ~=~ 0\right\}
  \lleq{bp-families}
  where
  \beq
  p_\rmBP(z_i) ~=~ \sum_{i=0}^{n+1} z_i^{d_i}
  \eeq
  is the weighted Fermat polynomial of degree $d$, with integral $d_i=d/w_i$.
  The monomials $z^J = \prod_{i=0}^{n+1}z_i^{j_i}$, with
  $J=(j_0,...,j_{n+1})$ for $j_i\in \mathN$ such that
  $\sum_{i=0}^{n+1} j_iw_i = d$ define the deformations.
  We denote the set of all deformation vectors $J$ that appear in the polynomial by
  $D \subset \mathZ^{n+2}$.

 In the context of this framework we analyze two separate but ultimately related
 issues. First, we generalize our previous results concerning the string theoretic
 interpretation of certain types of geometric modular forms from weighted Fermat
 hypersurfaces to families of Calabi-Yau varieties. Second, we use these generalized
 modularity results to investigate the conformal field theoretic structure of
 particular types of singular Calabi-Yau spaces.

The first step in this program involves the computation of geometric
$L-$series defined via motives $M_{\Om}(X(\psi_D))$ constructed from
the families $X(\psi_D)$ defined by (\ref{bp-families}). Our
construction of the motives of general families of
  $n-$dimensional weighted projective hypersurfaces involves the deformation
 of the $\Om-$motive defined at the Brieskorn-Pham point by the deformation
 monomials $z^J$, $J\in D$.  For any vector $J$ we define an
 action on the $\Om-$motive $M_\Om(X(0))$ of the Brieskorn-Pham variety $X(0)$, which
 we denote here by $\rho(J)M_\Om(X(0))$. The deformed motive of the family defined
 by $D$ is then defined as the sum of all contributions spanned by the action of the
 deformation vectors on the $\Om-$motive of the Brieskorn-Pham fiber
  \beq
   M_\Om(X(\psi_D)) = \bigoplus_{J\in D}~ \rho(J) M_\Om(X(0)).
  \lleq{deformed-omega-motive}
This construction will be made precise in Section 3.

For smooth fibers the motives $M_\Om(X(\psi_D))$ defined by
(\ref{deformed-omega-motive}) have in general higher rank and
therefore are not modular in terms the congruence subgroups of the
modular group $\rmSL(2,\mathZ)$. The Langlands program suggests
that such higher rank motives instead lead to automorphic forms
associated to higher rank groups. We will see later in this paper
that for singular fibers of the families of weighted projective
hypersurfaces the motives (\ref{deformed-omega-motive}) show
interesting degeneration behavior. As a result, families of
motives that are of high rank in the generic smooth fibers can
become modular for singular Calabi-Yau fibers. This phenomenon can
be used to investigate the modular structure of such singular
configurations. In the past, most of the work on singular string
compactifications has focused on conifolds, varieties with nodes.
In particular, the D-brane analysis by Greene, Strominger and
Morrison \cite{s95, gms95} of the geometric transitions introduced
in \cite{cdls88}, and analyzed in detail in \cite{gh88, cgh89,
cd90}, led to a paradigmatic interpretation.

Conifold singularities are not the only type of degeneration that
occur in moduli space, and the singularity types that we consider
in the present case are of more severe type, characterized by
Milnor numbers that are larger than one. For such more general
singularities a D-brane interpretation is lacking,  which raises
the question whether degenerate Calabi-Yau varieties of
non-conifold type admit a physical interpretation. For
Brieskorn-Pham varieties the underlying worldsheet theory is a
rational conformal field theory, and previous work has shown that
the modular forms derived for such spaces have a string theoretic
interpretation. The generalization of this string modularity
program away from rational theories has not been addressed
previously. The second purpose of this paper is to investigate the
conformal field theoretic properties of the singular fibers in
families of Calabi-Yau varieties. This leads to a surprising
result for the deformed motives. As mentioned above, for general
fibers of the Calabi-Yau families the deformed motive is of high
rank, and therefore not modular. The unexpected phenomenon
encountered in this work is that it can happen that the motivic
$L-$series of a non-Fermat type fiber in a family agrees with the
motivic $L-$series of a weighted Fermat variety. We will call such
pairs of models $L-$correlated. The existence of such a
$L-$correlated model shows that the deformed fiber inherits
information from the rational conformal field theory underlying
the Brieskorn-Pham manifold. This result makes it plausible that
singular configurations of such a type could serve as links, much
like conifolds, between different moduli spaces of smooth
configurations.

A second phenomenon we encounter in the present investigation is
that of dimensional reduction of the motives at the singular
fibers of the families. It happens that in some families of
varieties the modular forms of the deformed motives change their
weight at the singular fibers. The weight of these forms is lower
than the weight of the automorphic forms expected for the smooth
generic fiber of the family. Since the deformed $\Om-$motive
probes the dimension of the ambient space, and is not induced by
lower-dimensional subvarieties, this suggests that phase
transitions between Calabi-Yau varieties possibly involve motives
that effectively have different dimensions. Even in these extreme
 cases the modular forms that emerge contain information coming
 from the structure of the underlying worldsheet.
 We will call this phenomenon motivic dimensional transmutation.

 We exemplify our strategy with a number of one- and two-dimensional families
 of the type
  \beq
  X_n^d(\psi_1,\psi_2)
  ~=~ \left\{(z_0:\cdots z_{n+1}) \in \mathP_{(w_0,...,w_{n+1})} ~{\Big |}~
       p_\rmBP(z_i) - 2\psi_1 z_{n-2}^{a}z_{n-1}^{a}
    - 2\psi_2 z_n^{b}z_{n+1}^{b}=0\right\},
 \lleq{2-parameter-family-class}
 where we have simplified the notation for the deformation parameter for this
 particular case.
 For the special subclass of K3 surfaces these families extend the two-parameter family of
 quartic surfaces obtained for $d_i=4~ \forall i$, which was considered in depth in a
 different context and with different methods by Wendland \cite{kw04}.
 In Section 2 we describe the general motivic framework and in Section 3
 we detail  our construction of deformed $\Om-$motives and their $L-$functions.
 In Section 4 we compute these $L-$functions for families with up to two parameters
 of the type (\ref{2-parameter-family-class}) for K3 surfaces and Calabi-Yau threefolds.
 In Section 5 we focus on singular fibers of non-conifold type and present some
 modularity results, and in Section 6 we consider the conformal field theoretic aspects
 of the resulting modular forms. In Section 7 we illustrate the
 phenomenon of motivic dimensional transmutation for K3 surfaces and
 Calabi-Yau threefolds, and we conclude in Section 8.

\vskip .2truein

\section{Zeta functions and Weil conjectures for varieties and motives}

The Weil conjectures, proven by Dwork, Grothendieck and Deligne,
give a characterization of the congruence zeta function in terms
of the cohomology groups of the varieties, thereby turning a pure
counting function into an object that contains topological
information. In the past, conformal field theoretic applications
of arithmetic geometry have focussed on smooth projective
manifolds, or varieties in weighted projective spaces. In the
present paper we extend these applications to the case of more
general varieties, allowing for singular spaces. We therefore
briefly describe the behavior of the underlying arithmetic tools
in such a generalized framework.

\subsection{Arithmetic of general varieties}

A useful starting point for any discussion of motives is the
cohomological interpretation of the zeta function, conjectured
originally by Weil \cite{w49}, and proven by Grothendieck
\cite{g65}.  This cohomological interpretation amounts to a
factorization of this function which holds for any
 variety\fnote{1}{A variety is defined here as a separated scheme of
 finite type.}. Let $X$ be a variety of complex dimension $n$
 over a finite field $\mathF_q$, where $q=p^r$ for primes $p$
  and $r\in \mathN$. The Artin zeta function over $\mathF_q$, defined as
 \beq
  Z(X/\mathF_q,t) = \exp\left(\sum_{r\geq 1}
  \frac{N_{r,q}(X)}{r}t^r\right),
 \eeq
 factors as follows
 \beq
 Z(X/\mathF_q,t) = \frac{\prod_{i=0}^{n-1} \cP_q^{2i+1}(X,t)}{
                        \prod_{j=0}^{n} \cP_q^{2j}(X,t)},
 \eeq
 where $\cP_q^i(X,t)$ are polynomials whose  degree is given by the
 Betti numbers $b^i(X)=\rmdim~H^i(X)$ of the variety $X$
 \beq
 \rmdeg~\cP_q^i(X,t) ~=~ b^i(X).
 \eeq
 Consider the factorization of the polynomials
 \beq
   \cP_p^i(X,t) ~=~ \prod_j (1-\g_j^i(p)t).
 \eeq
 The generalized Riemann hypothesis is a characterization of the
 complex numbers $\g_j^i(p)$. The precise form of this hypothesis depends
 on the detailed structure of the variety. For smooth projective
 manifolds it was shown by Deligne \cite{d74} that the numbers
 $\g_j^i(p)$ are algebraic and satisfy
 \beq
 |\g_j^i(p)| = p^{i/2}.
 \eeq
 For general varieties which admit singularities a weaker constraint applies, as proven in
 \cite{d80}
 \beq
 |\g_j^i(p)| \leq p^{(i-\ell)/2},
 \eeq
for some $\ell \geq 0$. Here $\ell$ depends on the type of
singularity of the variety. We will describe several examples with
varying $\ell$ in this paper.

\subsection{Pure and mixed motives}

The notion of a motive of a variety was formulated by Grothendieck
as a means to identify the essential underlying geometric
structure that supports the various cohomology theories of a
variety. The goal is to isolate the fundamental irreducible
building blocks of arbitrary varieties, analogous to the idea of
fundamental particles in physics. In this picture, varieties are
viewed as constructed out of fundamental `geometric particles',
with the same motives appearing in completely different manifolds,
much like all atoms are built with a few particles. While the
concept of motives in its most general formulation has remained
somewhat elusive, in the present paper we build on the framework
of Grothendieck's effective motives with Tate twists. These are
defined via projectors $\g$ in the ring of correspondences, which
are multi-valued maps between varieties. A Grothendieck type
motive is therefore defined as a triplet
 \beq
  M~=~(X,\g,m),
 \eeq
  where $\g^2=\g$, and $m\in \mathZ$ denotes the Tate twist.
  The special class of motives with $m=0$ are called effective motives.
  Implicit in this definition is the choice of an
  equivalence relation that defines the correspondences.

 The correspondence $\g$ induces a projection on the cohomology,
 denoted by the same symbol, $\g H^*(X)$. This makes it possible to think of the
 motive $M$ in terms of its cohomological representation $H(M)$ for various types of
 cohomology groups $H^*(X)$ associated to the variety, i.e. $H(M) \subset H^*(X)$.
 Pure motives are distinguished by the fact that there exists an integer $w$,
 the weight $\rmwt(M)$ of the motive, such that the cohomological representative of the motive
 $M=(X,\g,m)$ is given by the projection $\g$ acting on the cohomology of degree $n$,
 where $\rmwt=n-2m$. Hence we have
 \beq
   H(M) ~=~ \g H^n(X)(m).
 \eeq
 It follows that the zeta function of an effective motive of weight $w$ defines a polynomial
 \beq
 Z(M/\mathF_q,t) ~=~ \cP_q(M,t)^{(-1)^{n+1}}
 \eeq
 that divides $\cP_q^n(X,t)$. The degree of this polynomial is determined
 by the rank of the motive, which can be defined in terms of $H(M)$ as
 \beq
 \rmrk~M ~=~ \rmdim~H(M)
 \eeq
 in $H^*(X)$. The $L-$function of a Grothendieck type motive can then be defined
 as the product over all primes with $t=p^{-s}$
 \beq
 L(M,s) ~=~ \prod_p \frac{1}{\cP_p(M,p^{-s})},
 \eeq
 where the detailed structure of the polynomials $\cP_p(M,p^{-s})$
 depends on whether the primes are good or bad. More details about the concept of pure
 motives can be found in \cite{rs08}. Physical applications of this notion
 in the context of transversal higher-dimensional hypersurfaces in weighted
 projective spaces are described in \cite{rs06,rs08,kls08}.
 A more advanced treatment can be found in \cite{jks94}.

A general formulation of mixed motives is not known at present. It
is expected that associated to each mixed motive $M$ is an
ascending weight grading $W_nM$, with $W_nM=0$ for $n<<0$ and
$W_nM=M$ for $n>>0$, such that the resulting quotients
 $\rmGr_n^wM = W_nM/W_{n-1}M$ are pure. A motive of weight $w$ is pure if
 $\rmGr_n^wM =0$ for $n\neq w$. The degree of the polynomials
 $\cP_p(M,t)$ of mixed motives is no longer determined by the simple
 relation that describes pure motives.

  \vskip .2truein

\section{The $\Om-$motive of Calabi-Yau families}

To define a Grothendieck motive means to define a projection $\g$.
In the present section we construct such $\g$ for general families
of Calabi-Yau hypersurfaces that contain a Brieskorn-Pham point in
their moduli space. Our idea is to consider projections that are
constructed from the deformations of the $\Om-$motives defined for
Brieskorn-Pham type hypersurfaces in \cite{rs06}, and for more
general varieties in \cite{rs08}. We denote the $\Om-$motive of
the weighted Fermat hypersurface $X(0)$ by $M_\Om(X(0))$. To
explain the construction of the motive of the family $X(\psi_D)$,
 with $\psi_J$ for $J\in D$ denoting the vector of moduli parametrizing the different
 deformations, as in (\ref{deformed-omega-motive}), it is necessary to construct
 an action $\rho_{J}$ induced by the deformation $z^J$ in the $J-$direction.
 The realization of this action depends on the particular representation
 of the motive $M_{\Om}(X(0))$ for the Brieskorn-Pham fiber. In the
 following, we briefly review the specialization of the $\Om-$motive to Brieskorn-Pham fibers
 \cite{rs08} and then define the action of the deformations on this motive.

With the ingredients introduced in the previous section one can
define a universal type of motive characteristic for Calabi-Yau
varieties as follows \cite{rs06,rs08}.

{\bf Definition.}~{\it Let $X$ be a Calabi-Yau manifold of
complex dimension $n$, $n\in \mathN$. Define the number field
$K_X$ as the extension of $\mathQ$ given by the
 algebraic numbers $\g_j^n$ defined by the factorization of the polynomial
 $\cP_p^n(t)$ associated to its intermediate cohomology,
 $K_X=\mathQ(\{\g_i^n(p)\})$. The cohomological representation of the $\Om-$motive
 $M_\Om(X)$ is defined as the orbit of the holomorphic $n-$form of $X$ with
 respect to the Galois group $\rmGal(K_X/\mathQ)$.}

 The notion of the $\Om-$motive can also be applied to the more
 general class of manifolds of special Fano type, which includes Calabi-Yau
 varieties as a special case. These spaces were originally introduced in the context
 of mirror symmetry in \cite{rs92,cdp93,rs94}, and have been
 discussed in the context of arithmetic mirror symmetry.
 It was shown in ref. \cite{kls08} that the modular forms of certain rigid Calabi-Yau
 spaces agree with the modular forms of their mirrors of special
 Fano type.

 Consider now the general class of Calabi-Yau hypersurface
 families defined in (\ref{bp-families}).
 In the context of these varieties the $\Om-$motive
 can be described concretely as the motive that is obtained from the
 action of the moduli on the $\Om-$motive of the Brieskorn-Pham fiber of the
family. In this framework an explicit formulation of the deformed
motive can be given as follows. Define for any weighted
hypersurface family $X_n^d(\psi_D)$ of dimension $n$ and degree
$d$ embedded in weigted projective space
$\mathP_{(w_0,...,w_{n+1})}$ the set of vectors
   \beq
  \cU = \left\{u\in \mathZ_{\geq 0}^{n+2}~{\Big |}~
     0\leq u_i \leq d_i-1 ~{\rm and}~ d{\Big |}\sum_i
     w_iu_i\right\},
 \lleq{u-set}
 where $d,w_i,d_i$ are as in eq. (\ref{bp-families}).
 To each vector $u\in \cU$ with $u_i\neq 0, \forall i$ we associate a
 monomial via $z^{u-1} = \prod_i z_i^{u_i-1}$. These monomials
 parametrize (part of) the cohomology of the fibers of the family.
 The vector $(1,1,\dots ,1)$, denoted by $u_\Om$ in the following,
 is distinguished by the fact that it represents the holomorphic
 $n-$form on the variety $X_n$.

 In terms of the vector $u_\Om$ defined above, the Brieskorn-Pham
 type $\Om-$motive can be parametrized by the Galois orbit
 \beq
  \cO_{\Om(0)} ~=~ \bigoplus_{g\in \rmGal(K_X/\mathQ)} \rho(g)~u_\Om,
 \lleq{omega-motive-orbit}
 where $\rho$ is the representation of $\rmGal(K_X/\mathQ)$ on
 $\cU$. Since $K_X$ is a cyclotomic field, the elements $g_n \in
 \rmGal(K_X/\mathQ)$ are defined on the generator $\xi$ of the field $K_X$
  as $g_n \xi = \xi^n$. The action $\rho(g_n)$ on $u$ then is defined as
 $\rho(g_n)u ~=~ nu(\rmmod~d)$.

 We construct the family of motives over the moduli space given by $(\psi_D)$
 via the orbit of $u-$vectors that act on the BP type motive.
 A general combination of monomial deformations is determined by a
 subset of $u-$vectors, denoted in eq. (\ref{bp-families})
  by $J \in D\subset \cU$.
 For any family of varieties defined by $D$ the deformed motive
 $M(X(\psi_D))$ is defined by the orbit that is obtained by considering
 \beq
 \cO_{\Om(\psi_D)} := \bigoplus_{J\in D}~r(J)~\cO_{\Om(0)},
 \lleq{deformed-omega-orbit}
 where the representation $r(J)$ on the $u-$vectors defining the $\Om-$motive via
 the orbit $\cO_{\Om(0)}$ is the additive action.
 The orbit $\cO_{\Om(\psi_D)}$ is to be viewed as a cohomological realization
 of the motive $M_\Om(X(\psi_D))$ introduced in eq. (\ref{deformed-omega-motive}).

 In order to determine the $L-$functions $L(M_\Om(X(\psi_D),s)$ of
 the motives $M_\Om(X(\psi_D))$ generated by the orbits $\cO_{\Om(\psi_D)}$, it is
 useful to first consider the cardinalities of the varieties in terms
 of Gauss sums. This is what we turn to in the next section.

\vskip .2truein

\section{Motivic $L-$functions for weighted hypersurface families}

In this section we apply our general construction to the special
case of families of Calabi-Yau hypersurfaces in weighted
projective spaces.

\subsection{Cardinalities}

For a complete motivic analysis of a variety, it is necessary to
have the complete cardinality structure of the variety under
control. There are several ways to compute this, using either
$p-$adic or complex methods. Both lend themselves equally well to
explicit computations. We find that the complex method is more
transparent as far as the motivic structure of the varieties is
concerned, while $p-$adic methods have traditionally been
pervasive in arithmetic geometry, ever since Dwork's proof of the
rationality of the zeta function (recent applications in physics
can be found in \cite{cov00, cov03, co07, k04, k06}). For
Brieskorn-Pham varieties complex techniques originated with Gauss
in the context of the cubic elliptic curve, and the special case
of the canonical family of Fermat hypersurfaces was considered by
Koblitz \cite{k83}.

The main ingredients in the computation of the cardinalities are
Gauss sums $G_{n,p}$ that are determined by two characters, the
multiplicative character $\chi_p$ and the additive character
$\Psi_p$, both associated to the finite field $\mathF_p$.
 The multiplicative character is given by the map
 \beq
  \chi_p:~~\mathF_p^\times ~\lra \mu_{p-1},
 \eeq
 defined by $ \chi_p(u) = \xi_{p-1}^m$,
 where $\xi_n=e^{2\pi i/n}$ and $m$ is determined by the generator
 $g\in \mathF_p^\times$ via $u=g^m$. The additive character
 \beq
 \Psi_p:~~\mathF_p ~\lra ~\mu_p,
 \eeq
 can be defined by $\Psi_p(u) = \xi_p^u$.

 The two characters $\chi_p$ and $\Psi_p$ can be lifted to characters
 on the degree $r$ extension $\mathF_q$ where $q=p^r$ for primes $p$ and
  $r\in \mathN$, via the norm and the trace map, respectively. The
 norm map is defined for $\mathF_q$ as
 \beq
  \rmN_r:~\mathF_q ~\lra ~\mathF_p
 \eeq
 where
 \beq
 \rmN_r(v) ~=~ v\cdot v^p \cdots v^{p^{r-1}}.
 \eeq
 The trace map is given by
 \beq
 \rmTr_r:~\mathF_q  ~\lra ~ \mathF_p
 \eeq
 with
 \beq
 \rmTr_r(v) ~=~ v + v^p +
                 v^{p^2} + \cdots + v^{p^{r-1}}.
 \eeq
 The lifts of the characters $\chi_p$ and $\Psi_p$ are then
 defined as the compositions
 \beq
 \chi_q(v) ~=~ \chi_p \circ \rmN_r,~~~~
 \Psi_q(v) ~=~ \Psi_p \circ \rmTr_r.
 \eeq

 The Gauss sum $G_{n,q}$ is defined in terms of the characters $\chi_q, \Psi_q$
  as
 \beq
 G_{n,q} ~=~ \sum_{v\in \mathF_q^\times} \chi_q(v)^n \Psi_q(v).
 \eeq
 In the context of the $p-$adic approach, the character $\Psi_p$ is
 replaced by the Dwork character, and the multiplicative character
 $\chi_p$ is replaced by the Teichm\"uller character (see e.g.
 \cite{cov00}).

The additive character satisfies the vanishing relation
 \beq
 \sum_{x\in \mathF_q} \Psi_q(yx)
  ~=~  \left\{\begin{tabular}{c l}
              $q$  &if $y=0$ \\
               0   &if $y\in \mathF_q^\times$ \tabroom \\
      \end{tabular}
  \right\}
 \eeq
 and therefore lends itself to the computation of cardinalities of polynomial varieties.

 An efficient strategy for deriving the cardinalities of projective varieties and their motives
 is to first consider the multiplicative cardinalities of the affine varieties
 $X(\psi_D)_\rmaff$, denoted here by $N_q^\times(X(\psi_D)_\rmaff)$.
 The projective cardinalities can be determined by first computing the affine
 cardinalities $N_q(X(\psi_D)_\rmaff)$ via an iterative procedure that takes into
 account the multiplicative affine cardinalities of lower-dimensional strata,
 and by implementing the projective equivalence relation. To this effect, we define
 the subvarieties $X_{n-j}(\psi_D)$ of $X_n(\psi_D)$ via the hyperplane
 intersections
 \beq
 H_{i_1,...,i_j} ~=~ \{x_{i_1}=0\} \cap \cdots \cap \{x_{i_j}=0\}
  \cap \mathP_{(w_0,...,w_{n+1})}
 \eeq
 as
 \beq
 X_{n-j}(\psi_D) ~=~ X_n(\psi_D) \cap H_{i_1,...,i_j}.
 \lleq{lower-dim-subvars}
 Collecting all the contributions of the multiplicative affine cardinalities of the subvarieties
 (\ref{lower-dim-subvars}) leads to the affine cardinalities
 \beq
  N_q(X(\psi_D)_\rmaff) ~=~ \sum_{j=0}^{n-1}
  N_q^\times(X_{n-j}(\psi_D)_\rmaff).
 \eeq
 Projectivizing the affine cardinalities then leads to
 cardinalities of the projective varieties
  \beq
  N_q(X_n(\psi_D)) ~=~ \frac{1}{q-1}(N_q(X_n(\psi_D)_\rmaff)-1).
  \eeq
The motives that we consider are those of the projective
varieties.

 The starting point for the derivation of the affine multiplicative cardinalities
 is their representation in terms of the additive character
 \beq
 N_q^\times(X_n^d(\psi_D)_\rmaff)
  ~=~ \frac{1}{q} \sum_{y\in \mathF_q} \sum_{z_i\in \mathF_q^\times}
   \Psi_q(yP(z_i,\psi_D)),
 \eeq
 where $P(z_i,\psi_D)$ is the polynomial in eq.
 (\ref{bp-families}).
In order to perform the $\mathF_q-$sums, it is useful to factorize
$P(z_i,\psi_D)$ as much as possible. This can be achieved by trading
the additive character for the multiplicative character by inverting
the Gauss sums
 \beq
 \Psi_q(x) ~=~ \frac{1}{q-1} \sum_{m=0}^{q-2} G_{-m,q}\chi_q(x)^m.
 \eeq
 The final input needed in the computation is the relation
 \beq
 \sum_{x\in \mathF_q^\times} \chi_q(x)^m
  ~=~ \left\{\begin{tabular}{c l}
      $(q-1)$   &if $(q-1)|m$ \\
       0    &otherwise \tabroom \\
      \end{tabular}
  \right\}.
 \eeq

With these ingredients we obtain explicit results for the
cardinalities of families of weighted projective hypersurfaces
 $X_n \subset \mathP_{(w_0,...,w_{n+1})}$ as well as the motives defined
 above. In what follows, we introduce for $q$ with $d|(q-1)$ the integer
 $k=(q-1)/d$. For families of varieties the cardinality formulae depend on the
detailed form  of the deformations. In the following, we consider
certain classes of one- and two-parameter families of weighted
hypersurfaces in arbitrary dimensions such that the family
contains a Brieskorn-Pham type fiber. It is useful to introduce
the following Gauss sum products for the zero-, one-, and
 two-parameter families
 \bea
   \mathG_q^n(u)
      &=& \prod_{i=0}^{n+1}  G_{-w_iu_ik,q}  \nn \\
  \mathG_q^n(u,\psi)
      &=& \mathG_q^{n-2}(\uu)  \sum_{\ell=0}^{\frac{d}{2}k-1}
           G_{2\ell,q} \chi_q(-2\psi)^{-2\ell}
         \left(\prod_{j=n}^{n+1} G_{-(\ell+w_ju_jk),q}\right) \nn  \\
   \mathG_q^n(u,\psi_1,\psi_2)
     &=& \mathG_q^{n-3}(\uu) \sum_{\ell_1,\ell_2=0}^{\frac{d}{2}k-1}
          \mathG_{2\ell_1,q} \mathG_{2\ell_2,q}
          \chi_q(-2\psi_1)^{-2\ell_1} \chi_q(-2\psi_2)^{-2\ell_2} \nn \\
     &&~~~~~~~~~~~\cdot \left(\prod_{i=n-2}^{n-1} G_{-(\ell_1+w_iu_ik),q} \right)
          \left(\prod_{j=n}^{n+1} G_{-(\ell_2+w_ju_jk),q}\right).
  \llea{gauss-sum-products}
 Here $\uu$ denotes the truncated $u-$vector of the reduced Gauss
 sum product. For the one-parameter family $\uu=(u_0,...,u_n)$,
 while for the two-parameter case $\uu=(u_0,...,u_{n-1})$.

We find that the cardinalities or our varieties can be written in
terms of the Gauss sum polynomials of eq.
(\ref{gauss-sum-products}) in the following expressions, amenable
to explicit computations.

 {\bf Result 1.}~ {\it The multiplicative affine cardinalities of
  Brieskorn-Pham type hypersurfaces of dimension $n$ and degree $d$
  for prime powers $q$ such that $d|(q-1)$ are given by}
 \beq
   N_q^\times(X_n(0)_\rmaff)
  ~=~ \frac{(q-1)^{n+2}}{q} + \frac{(q-1)}{q} \sum_{u\in \cU} \mathG_q^n(u),
 \lleq{bp-cards-gauss}

For the  multiplicative affine cardinalities of the one- and
two-parameter families the remaining Gauss sum products of
(\ref{gauss-sum-products}) enter.

{\bf Result 2.}~{\it Define the one-parameter family $X_n^d(\psi)$
by}
 \beq
  X_n^d(\psi) ~=~ \left\{(z_0: \cdots : z_{n+1}) \in  \mathP_{(w_0,...,w_{n+1})} ~{\Big |}~
     p_\rmBP(z_i) - 2\psi z_n^az_{n+1}^a=0\right\}.
  \eeq
 {\it The multiplicative affine cardinalities for prime powers $q$ such that
  $d|(q-1)$ are then given by}
 \beq
   N_q^\times(X_n^d(\psi)_\rmaff)
   ~=~ \frac{(q-1)^{n+2}}{q}
    + \frac{1}{q} \sum_{u\in \cU} \mathG_q^n(u,\psi).
  \eeq
 {\it For the two-parameter families}
 \beq
  X_n^d(\psi_1,\psi_2)
  = \left\{(z_0:\cdots : z_{n+1})\in \mathP_{(w_0,...,w_{n+1})} ~{\Big |}~
   p_\rmBP(z_i) - 2\psi_1 z_{n-2}^az_{n-1}^a
                        - 2\psi_2 z_n^bz_{n+1}^b=0\right\}
  \eeq
  {\it the cardinalities  are given by}
 \beq
  N_q^\times(X_n^d(\psi_1,\psi_2)_\rmaff)
   ~=~ \frac{(q-1)^{n+2}}{q} + \frac{1}{q(q-1)} \sum_{u \in \cU}
   \mathG_q^n(u,\psi_1,\psi_2).
  \eeq
  {\it Here the Gauss sum products $\mathG_q^m(u,\psi)$ and
    $\mathG_q^m(u,\psi_1,\psi_2)$ are defined as in
    (\ref{gauss-sum-products}).}

The cardinalities for other primes can be determined with the same
techniques. In general, more primes than the ones considered in the
second result are necessary to determine the modular forms uniquely,
not only up to twists. For the hypersurfaces considered in this
paper, the primes of the type $p\equiv 1(\rmmod~d/2))$ are
sufficient to uniquely determine the motivic modular form.

\subsection{$L-$functions of the motives}

By projectivizing the cardinality results obtained above we define
the contributions of the $\Om-$motive to the cardinalities for the
Brieskorn-Pham varieties as
 \beq
 M_\Om(X_n^d(0)):~~~ a_p~ =~ \frac{(-1)^{n-1}}{p} ~\sum_{u\in
 \cO_{\Om(0)}} \mathG_p^n(u).
 \lleq{bp-motivic-cards}

 For the fibers of a family with several parameters we denote by $|D|$ the
 dimension of the family. For deformed motives that receive no contributions
 from lower dimensional strata in the transition from multiplicative affine to
 projective cardinalities, we define the motivic cardinalities as
 \beq
 M_\Om(X_n^d(\psi_D)):~~~
 a_p(\psi_D) ~=~ \frac{(-1)^{n-1}}{p(p-1)^{|D|}}
   ~\sum_{u\in \cO_{\Om(\psi_D)}} \mathG_p^n(u,\psi_D).
 \lleq{family-motive-cards}
 With the convention that $\mathG_p(u,0) = \mathG_p(u)$, for $|D|=0$
 (the case of no deformation), this expression formally reduces to
 (\ref{bp-motivic-cards}) for Brieskorn-Pham motives.

 If the zeta function polynomials associated to the motive do not
 factorize, the coefficients $a_p$ computed in this way determine the
 coefficients of the $L-$function of the $\Om-$motive
  \beq
   L_\Om(X(\psi_D),s) ~=~ L(M_\Om(X(\psi_D)),s) ~=~ \sum_n
   \frac{a_n}{n^s}.
  \eeq
 In the case the zeta function factorizes corresponding modifications have to
 be implemented, depending on the type of factorization.

In this paper we are mainly interested in the highest weight
motives because they are most characteristic of the variety. The
strategy described above for computing the $L-$function of these
motives, however, applies to all motives defined by our
deformation construction. The basic structure is that the
intermediate cohomology group $H^n(X_n)$ of the variety $X_n$
decomposes into orbits $\cO_i$, each of which defines a motive
$M_i$. The $L-$functions of these motives $M_i$ then lead to lower
weight modular forms via Tate twists.

In the remainder of this paper, we consider several examples of K3
and threefold families and some of their physical implications.

\section{Modular phases of CY varieties}

In this section we discuss in detail several examples provided by
fibers of two 1-dimensional K3 families and two 2-dimensional
families of Calabi-Yau threefolds. Furthermore, we list results for
additional K3 surfaces and threefolds in Tables 2 and 3. Our
discussion involves a number of different modular forms of different
weights and levels, hence it is useful to collect these objects for
later reference in Table 1\fnote{1}{Resources for modular forms are
William Stein's websites, as well as the tables of Meyer \cite{cm05}
and Sch\"utt \cite{ms06}.}. Some of these modular forms can be
expressed in closed form in terms of the Dedekind eta function and
the Eisenstein series. Denoting the Eisenstein field by $K_E =
\mathQ(\sqrt{-3})$, these function are given as
 \beq
 \eta(q)=q^{1/24}\prod_{i\geq 1} (1-q^n),
 ~~~~~~~\vartheta(q) = \sum_{z\in \cO_{K_E}} q^{\rmN z},
 \lleq{dedekind-eisenstein}
 respectively, where $q=e^{2\pi i \tau}$, with $\tau$ in the upper half
 plane, and $\rmN z$ denotes the norm of the algebraic integer
 $z\in \cO_{K_E}$.
 \begin{small}
 \begin{center}
 \begin{tabular}{r l}
  Modular form    &$q-$Expansion or closed form \\
  \hline
   $f_{2,27}(q)~=~\eta^2(q^3)\eta^2(q^9)$
                  &$=~ q - 2q^4 - q^7 + 5q^{13} + 4q^{16} - 7q^{19} - 5q^{25}
                       + 2q^{28} - 4q^{31} + 11q^{37} + \cdots $\tabroom \\

   $f_{2,32}(q)~=~\eta^2(q^4)\eta^2(q^8)$
                &$=~ q - 2q^5 - 3q^9 + 6q^{13} + 2q^{17} -q^{25} - 10q^{29} - 2q^{37} + 10q^{41}
                + \cdots $  \tabroom \\

   \hline

   $f_{3,16}(q)~=~\eta^6(q^4)$
                &$=~ q - 6q^5 + 9q^9 + 10q^{13}  - 30q^{17} + 11q^{25} + 42q^{29} - 70q^{37}
                   %+ 18q^{41}
                   +  \cdots $ \tabroom \\
   $f_{3,27}(q)~=~\eta^2(q^3)\eta^2(q^9)\vartheta(q^3)$
               &$=~ q + 4q^4 - 13q^7 - q^{13} + 16q^{16} + 11q^{19} + 25q^{25} - 52 q^{28}
                  - 46q^{31} % + 47q^{37}
                    +  \cdots $ \tabroom \\
   $f_{3,108}(q)$&$=~q + 11q^7 + 23q^{13} - 37q^{19} - 46q^{31} - 73q^{37} - 22q^{43}
                        + 47q^{61} %- 13q^{67}
                        + \cdots$
                         \tabroom \\
   \hline

   $f_{4,9}(q)~=~\eta^8(q)$
                 &$=~ q - 8q^4 + 20q^7  - 70q^{13} + 64q^{16} + 56 q^{19} - 125q^{25}
          - 160q^{28} % + 308q^{31} + 110q^{37}
          + \cdots$
                   \tabroom  \\
   $f_{4,32}(q)$&$=~ q + 22q^5 - 27q^9 - 18q^{13} - 94q^{17} + 359 q^{25} - 130q^{29} + 214q^{37}
                     %- 230q^{41}
                     +  \cdots$
                         \tabroom \\
   $f_{4,108\rmA}(q)$&$=~ q - 37q^7 - 19q^{13} - 163q^{19} -125q^{25} + 308q^{31} + 323q^{37}
                      % - 520q^{43} + 1026q^{49}
                      + \cdots$   \tabroom \\
   $f_{4,108\rmB}(q)$&$=~ q + 17q^7 + 89q^{13} + 107q^{19} -125q^{25} + 308q^{31} - 433q^{37}
                      %- 520q^{43} -54q^{49}
                    + \cdots$
                               \tabroom \\
  \hline
\end{tabular}
\end{center}
 \end{small}
 \centerline{{\bf Table 1.}~{\it Motivic modular form factors $f_{w,N}(q) \in S_w(\G_1(N))$
  of weight $w$ and level $N$.}}

Some of the  motivic forms identified below are twists of the forms listed in Table 1, involving
the Dirichlet characters defined by the Legendre symbol, denoted here by
 \beq
 \chi_n(p) ~=~ \left(\frac{n}{p}\right)
  ~=~ \left\{\begin{tabular}{r l}
         1  &if $x^2\equiv n$ for $x\in \mathF_p$ \\
         $-1$ &otherwise \tabroom \\
     \end{tabular}
     \right\}.
 \eeq
 The level of the twisted form $f\otimes \chi$ depends on the
 level $N$ of $f$, the conductor $C_\e$ of the character $\e$ of
 $f\in S_w(\G_0(N),\e)$, and the conductor $C_\chi$ of $\chi$.
 With $C_\e|N$ the level is given by
 \beq
 N(f\otimes \chi) ~=~ \rmlcm\{N,C_\e C_\chi, C_\chi^2\}.
 \eeq

The type of families we discuss in more detail are deformations of
the Brieskorn-Pham hypersurfaces considered in \cite{rs06,rs08}.
Specifically, for K3 sufaces, we consider the families defined by
 \bea
  X_2^{6\rmB}(\psi) &=& \left\{ z\in \mathP_{(2,2,1,1)}~{\Big |} ~
                          z_0^3 + z_1^3 + z_2^6 + z_3^6
                            - 2\psi z_2^3z_3^3  =0 \right\} \nn \\
  X_2^8(\psi) &=& \left\{ z\in \mathP_{(4,2,1,1)}~{\Big |} ~
                          z_0^4 + z_1^4 + \sum_{i=2}^3 z_i^8 - 2\psi
                          z_2^4z_3^4  =0 \right\},
 \llea{k3-examples}
 while for Calabi-Yau threefolds we analyze in more detail the two-parameter families
 \bea
  X_3^6(\psi_1,\psi_2) &=& \left\{ z\in \mathP_{(2,1,1,1,1)}~{\Big |} ~
                          z_0^3 + \sum_{i=1}^4 z_i^6
                          -2\psi_1z_1^3z_2^3 - 2\psi_2 z_3^3z_4^3  =0 \right\} \nn \\
   X_3^8(\psi_1,\psi_2) &=& \left\{ z\in \mathP_{(4,1,1,1,1)}~{\Big |} ~
                          z_0^2 + \sum_{i=1}^4 z_i^8
                          -2\psi_1z_1^4z_2^4 - 2\psi_2 z_3^4z_4^4  =0 \right\}.
  \llea{cy3-examples}
 The $\Om-$motives of the Brieskorn-Pham fibers at the origin of
 the families $X_2^{6\rmB}(\psi)$ and $X_3^6(\psi_1,\psi_2)$ were shown in
 \cite{rs06,rs08}
 to be modular  of weight 3 and 4, and levels $N=27$ and $N=108$, respectively.
 Modularity of the Brieskorn-Pham $L-$series follows from the fact that they are
 of complex multiplication type, i.e. they are purely number theoretic objects
 determined by Hecke characters $\Psi_H$. Such Hecke type $L-$series have been
 shown to be modular in Hecke's work \cite{h59}, hence it follows
 that the motives that give rise to these characters  are modular.
 The Hecke representation of the motivic modular forms of $X_2^{6\rmB}(0)$ and $X_3^6(0)$
 allows us to express the resulting modular forms in terms of Hecke
 indefinite modular forms that arise in the underlying conformal
 field theory \cite{rs06,rs08}.

 The rank of the motives of the smooth generic fibers of the above families
 $X_2^{6\rmB}(\psi)$ and $X_3^6(\psi_1,\psi_2)$ varies between four and eight,
 indicating that the degeneration of the zeta functions depends strongly on
 the type of singularity of the configuration.

\subsection{K3 surfaces}

In this subsection we discuss two one-dimensional families in
detail and summarize further results in Table 2.

 \underline{$X_2^{6\rmB}(\psi) \subset \mathP_{(2,2,1,1)}$}

 Our first example is the family of elliptic K3 surfaces $X_2^{6\rmB}(\psi)$
 of eq. (\ref{k3-examples}).
 The extremal Brieskorn-Pham fiber at $\psi=0$ has been analyzed in detail
 in \cite{rs06}, in combination with the two other extremal K3 surfaces of
  Brieskorn-Pham type, $X_2^4(0) \subset \mathP_3$ and
  $X_2^{6\rmA}(0) \subset \mathP_{(3,1,1,1)}$. To apply our
  construction,
 consider first the $\Om-$motive $M(X_2^{6\rmB}(0))$ of the Brieskorn-Pham fiber
 in this family $X_2^{6\rmB}(\psi)$. This motive is given by the Galois orbit of the
 holomorphic 2-form, as defined in (\ref{omega-motive-orbit}), and can be represented
 by the $u-$vector orbit as
 \beq
 M_\Om(X_2^{6\rmB}(0)):~~~(1,1,1,1) ~\oplus ~ (2,2,5,5).
 \eeq
The motive of the family $\psi\neq 0$ is obtained as the orbit
generated by the deformation vector $J=(0,0,3,3)$ acting on the
$\Om-$motive $M(X_2^{6\rmB}(0))$, as defined in
(\ref{deformed-omega-orbit}), leading to
 \beq
 M_\Om(X_2^{6\rmB}(\psi)):~~~ (1,1,1,1) ~\oplus ~ (1,1,4,4) ~\oplus ~(2,2,5,5) ~\oplus ~
 (2,2,2,2).
 \eeq
 For the $L-$series of the deformation motive $M_\Om(X_2^{6\rmB}(1))$
 we obtain from eq. (\ref{family-motive-cards})
  \beq
 L_\Om(X_2^{6\rmB}(1),s)
  \doteq 1 - \frac{13}{7^s} - \frac{1}{13^s} + \frac{11}{19^s}
      - \frac{46}{31^s} + \frac{47}{37^s} + \cdots
 \eeq
Here we ignore the bad primes at $p=2,3$.

Surprisingly, this $L-$function, given by the motive of a model
that describes the deformation of a Brieskorn-Pham variety, is
identical to the $L-$series of a different Brieskorn-Pham type K3
manifold, the surface $X_2^{6\rmA}(0) \subset \mathP_{(3,1,1,1)}$
of the one-parameter family $X_2^{6\rmA}(\psi)$. The $L-$series
$L_\Om(X_2^{6\rmA}(0),s)$ has been computed previously in
\cite{rs06} via Jacobi sums, but also can be obtained by using the
motivic cardinalities (\ref{bp-motivic-cards}) in combination with
the Gauss sums in eq. (\ref{gauss-sum-products}). These
computations lead to the identity
 \beq
  L_\Om(X_2^{6\rmB}(1),s) ~=~ L_\Om(X_2^{6\rmA}(0),s).
 \eeq
 This fact has a number of immediate consequences. First, the $L-$function
 of the deformed motive is a Hecke $L-$series because the $L-$series of the Brieskorn-Pham
 hypersurface is determined by the Hecke character associated to the
 Eisenstein field $K_E = \mathQ(\sqrt{-3})$, defined by
  $\psi_{\rm 6A}(\pfrak) = \pfrak$ for $\a_\pfrak \equiv
  1(\rmmod~3)$. It therefore
  follows from the work of Hecke that the corresponding $q-$series
 is modular. The result is the modular form $f_{3,27}(q)$ of weight 3 and level
 $N=27$, listed in Table 1,
 \beq
 f_\Om(X_2^{6\rmA}(0),q) ~=~
    \eta^2(q^3)\eta^2(q^9) \vartheta(q^3).
\eeq
 Here $\eta(q)$  and $\vartheta(q)$ are as
 defined in (\ref{dedekind-eisenstein}).

 The relation between the deformed singular fiber and the weighted
 Fermat fiber thus immediately implies
 that the deformed $\Om-$motive of $X_2^{6\rmB}(1)$ is modular,
 with the modular form given by that of the Brieskorn-Pham K3 surface
 \beq
   f_\Om(X_2^{6\rmB}(1),q) ~=~ f_\Om(X_2^{6\rmA}(0),q) ~\in ~ S_3(\G_1(27)).
 \eeq
 This result also shows that the string model of the underlying
 deformed singular conformal field theory can lead to precisely
 the same structure as an exactly solvable model corresponding
 to a smooth Brieskorn-Pham hypersurface, i.e. a rational theory.
 We will expand on this in Section 6, when we discuss the worldsheet
 interpretation of this result.

The fact that this K3 surface is elliptic suggests that it should
be possible to identify modular forms of weight two in the
singular fiber of the family $X_2^{6\rmB}(\psi)$. The generic
elliptic fiber in this family is a cubic curve embedded in the
projective plane $\mathP_2$. The elliptic motive in the family is
given by the orbit
 \beq
 M_\rmell(X_2^{6\rmB}(1)):~~~~
  (1,1,2,0) \oplus (1,1,5,3) \oplus (2,2,4,0) \oplus (2,2,1,3),
 \eeq
 and the associated $L-$function is given by
 \beq
  L(M_\rmell(X_2^{6\rmA}(1)),s)
   ~\doteq~ 1 - \frac{1}{7^s} + \frac{5}{13^s} - \frac{7}{19^s}
         - \frac{4}{31^s} + \frac{11}{37^s} +
         \cdots
 \eeq
 This leads to the unique modular form of weight two and level 27,
 given by $f_{2,27}(q) = \eta^2(q^3)\eta^2(q^9) \in
 S_2(\G_0(27))$. In a string theoretic context this form was
 discussed in \cite{su02}, where it was shown that this worldsheet
 modular form matches that of the elliptic Fermat curve
 \beq
 E^3 = \left\{(z_0:z_1:z_2)\in \mathP_2 ~{\Big |}~z_0^3+z_1^3+z_2^3=0\right\}.
 \lleq{cubic-fermat-curve}

 \underline{$X_2^8(\psi) \subset \mathP_{(4,2,1,1)}$}

The family $X_2^8(\psi)$ of octic K3 surfaces is elliptically fibered
with generic fibers that are smooth quartic curves in the weighted
projective plane $\mathP_{(2,1,1)}$. The $\Om-$motive at the
Brieskorn-Pham point $\psi=0$ is of rank four, and can be
represented as the Galois orbit
 \beq
 M_{\Om}(X_2^8(0)):~~~(1,1,1,1) \oplus (1,3,3,3) \oplus (1,1,5,5) \oplus
 (1,3,7,7).
 \eeq
 The deformed motive determined by the vector $J = (0,0,4,4)$
 does not change this motive, hence we have in this particular
 case
 $$
 M_{\Om}(X_2^8(\psi)) ~=~ M_{\Om}(X_2^8(0))
 $$
 in the representation (\ref{deformed-omega-motive}) and (\ref{deformed-omega-orbit}).

 For the $L-$series of the motive $M_\Om(X_2^8(1))$ we find with the
 one$-$parameter formula of eq. (\ref{gauss-sum-products})
 \beq
 L_\Om(X_2^8(1),s)
   ~\doteq~ 1 + \frac{6}{5^s} - \frac{10}{13^s} - \frac{30}{17^s}
        - \frac{42}{29^s} + \frac{70}{37^s} + \frac{18}{41^s}
      % - \frac{110}{73^s} - \frac{78}{89^s}
      + \cdots
 \eeq
 This motivic $L-$series is also identical to the twist of the $L-$function
 of a Fermat hypersurface, in this case the quartic K3 surface
 $X_2^4(0) \subset \mathP_3$, defined by
 \beq
 X_2^4(0) ~=~ \left\{(z_0:\cdots :z_3) \in \mathP_3 ~{\Big |}~
     \sum_{i=0}^4 z_i^4 = 0\right\}.
 \eeq
 The twist character is given by the quadratic character $\chi_2$ associated to the Gauss
 field $K_G = \mathQ(\sqrt{-1})$
 \beq
 L_\Om(X_2^8(1),s)~=~ L_\Om(X_2^{4}(0),s) \otimes \chi_2.
 \eeq
 The $L-$series of the quartic Fermat surface $X_2^{4}(0)$ is again a number
 theoretic object, determined by a Hecke character $\Psi_H$ associated
 to the Gauss field $K_G$
 \beq
 L_\Om(X_2^{4}(0),s)~=~ L(\Psi_H^2,s).
 \eeq
  Hence it follows again from Hecke's theory that this $L-$series is
 modular and that the modular form of the octic K3 surface is of CM-type,
 given by
 \beq
 f_\Om(X_2^8(1),q) ~=~ \eta^6(q^4) \otimes \chi_2 ~\in ~ S_3(\G_1(64)),
 \eeq
 with $\eta^6(q^4) \in S_3(\G_0(16),\chi_{-1})$.

 The string theoretic structure of the modular motive of the quartic K3
 surface $X_2^{4}(0)$ was discussed in \cite{rs06}. The underlying rational conformal field theory of
 $X_2^{4}(0)$ leads to Hecke indefinite modular forms $\Theta^k_{\ell,m}(\tau)$
 with $k=2$, and the  theta series $\Theta^2_{1,1}(q) = \eta(q)\eta(q^2)$
 completely determines the weight three modular form $\eta^6(q)$ via the symmetric
 square. Hence its twisted relative $f_\Om(X_2^8(1),q)$ inherits the same rational conformal field theoretic
 structure, modified only by the appearance of the Dirichlet character.

The elliptic fibration structure of this K3 surface implies that
there is a lower weight motive that provides a further modular
object in this model. The elliptic motive is given by
 \beq
 M_{\rmell}(X_2^8(1)):~~~(1,1,2,0) \oplus (1,1,6,4) \oplus (1,3,6,0) \oplus
 (1,3,2,4),
 \eeq
 which leads to the $L-$series
 \beq
 L(M_\rmell(X_2^8(1),s) ~=~ L(f_{2,32},s) \otimes \chi_2,
 \eeq
 where $f_{2,32}$ is the modular form of weight $w=2$ and level
 $N=32$ listed in Table 1. The modular form $f_{2,32} \otimes
 \chi_2$ is an element of $S_2(\G_0(64))$. Its $L-$series is known
 to arise from the elliptic curve
  \beq
   E^4 ~=~ \left\{(z_0:z_1:z_2)\in \mathP_{(1,1,2)}~{\Big |}~
      z_0^4 + z_1^4 +z_2^2 =0\right\},
 \eeq
 a curve in the configuration of curves that appear in the elliptic
 fibration \cite{ls04}.

\underline{More K3 results:}

Other K3 surfaces can be analyzed in the same way. In Table 2 we
summarize the discussion so far and add further results. The K3
family $X_2^{12}(\psi) \subset \mathP_{(6,4,1,1)}$ is also an
elliptic fibration, with the generic fiber given by degree six
elliptic curves in the weighted projective plane
$\mathP_{(1,2,3)}$. The $L-$series $L_\Om(X_2^{12}(1),s)$ of the
singular fiber of the family is again a Hecke $L-$series
associated to a Hecke character, and is therefore modular. The
resulting modular form admits complex multiplication and leads to
a twist of the level $N=27$ modular form $f_{3,27}(q)$ of weight 3
listed in Table 1. This form is thus again a symmetric square of a
weight two modular form, this time associated to the
Brieskorn-Pham elliptic curve embedded in $\mathP_{(1,2,3)}$. The
structure of this K3 surface is therefore analogous to that of the
degree six surface $X_2^6(1)$ considered in detail above, with the
difference that the generic elliptic fiber is now of degree six,
not of degree three.
 \begin{center}
\begin{tabular}{l r r}

K3 surface    &Modular form    \\
\hline

 $X_2^{6\rmA}(1) \subset \mathP_{(3,1,1,1)}$
             &$f_{3,108}\otimes \chi_3 \in S_3(\G_1(432))$
              \tabroom \\

 $X_2^{6\rmB}(1) \subset \mathP_{(2,2,1,1)}$
             &$f_{3,27} \in S_3(\G_1(27))$
              \tabroom \\

 $X_2^{8}(1) \subset \mathP_{(4,2,1,1)}$
             &$f_{3,16} \otimes \chi_2 \in S_3(\G_1(64))$ \tabroom \\

 $X_2^{12}(1) \subset \mathP_{(6,4,1,1)}$
            &$f_{3,27}\otimes \chi_3 \in S_3(\G_1(432))$ \tabroom \\
 \hline
 \end{tabular}
 \end{center}

\centerline{{\bf Table 2.}~{\it Modular fibers in 1-parameter K3
families.}}

\subsection{One and two-dimensional families of Calabi-Yau threefolds}

Consider two-dimensional Calabi-Yau threefold families of the type
  \beq
 X_3^d(\psi_1,\psi_2)
  ~=~ \left\{(z_0:\cdots z_{4})\in
  \mathP_{(w_0,...,w_4)} ~{\Big |}~
    \sum_{i=0}^{2} z_i^{d_i} - 2\psi_1 z_1^az_2^a
   - 2\psi_2 z_2^bz_{3}^b=0\right\},
 \lleq{special-class}
 with $d=3,4$, $a=d/2w_1, b=d/2w_3$, and
 $(w_0,...,w_4) \in \{(2,1,1,1,1), (4,1,1,1,1)\}$.
 The rank of the motive for the generic fibers is
 eight in both cases, but for the singular fibers at $(\psi_1,\psi_2)=(1,1)$
 the $L-$function of the motive degenerates and becomes modular.
 We exemplify our general construction with these two cases in turn.

\vskip .1truein

\underline{$X_3^6(\psi_1,\psi_2) \in \mathP_{(2,1,1,1,1)}$}

This two-parameter family extends the Brieskorn-Pham variety
$X_3^6(0)$ considered in \cite{rs08}, where it was shown that the
 $L-$series of the $\Om-$motive
 \beq
  M_{\Om}(X_3^6(0)):~~~~ (1,1,1,1,1) ~\oplus ~ (2,5,5,5,5),
 \eeq
 is modular, arising from the modular form
  \beq
  f_\Om(X_3^6(0),q) ~=~ f_{4,108A}(q) ~\in ~S_4(\G_0(108)),
  \eeq
 where $f_{4,108A}$ is identified by its expansion in Table 1.

 The deformed $\Om-$motive of the two-parameter family $X_3^6(\psi_1,\psi_2)$
 is parametrized by the orbit
 \beq
 M_{\Om}(X_3^6(\psi_1,\psi_2)):~~~~
  \langle
  M_{\Om}(X_3^6(0),~J^1=(0,0,0,3,3),~J^2=(0,3,3,0,0)\rangle,
 % (1,1,1,1,1) ~\oplus ~(1,1,1,4,4) ~\oplus ~(1,4,4,4,4) ~\oplus ~
 % (2,5,5,2,2) ~\oplus ~(2,2,2,5,5) ~\oplus ~(2,2,2,2,2)
 \eeq
 generated by the two deformation vectors.
 While generically this motive is of rank eight, it degenerates
 for the singular fibers. For the fiber at
 $(\psi_1,\psi_2)=(1,1)$,
 we find for the motivic $L-$function
 \beq
 L_\Om(X_3^6(1,1),s)
  = 1 + \frac{20}{7^s} - \frac{70}{13^s} + \frac{56}{19^s}
      + \frac{308}{31^s}  + \frac{110}{37^s} + \cdots
 \eeq
 Here the bad primes are $p=2,3$.
 This $L-$series has complex multiplication and is determined by the Hecke $L-$series of a Hecke
 Gr\"o\ss encharacter $\psi_H$ associated to the Eisenstein field
 $\mathQ(\sqrt{-3})$
  \beq
  L_\Om(X_3^6(1,1),s) ~=~ L(\psi_H^3,s).
  \eeq
   Hecke's theory therefore implies that this $L-$series is modular.
 The associated modular form admits a closed expression in terms of the
 Dedekind eta function
 \beq
 f_{\Om}(X_3^6(1,1),q) ~=~ \eta(q^3)^8 ~\in ~S_4(\G_0(9)),
 \eeq
 denoted by $f_{4,9}(q)$ in Table 1.
 This CM form is furthermore string theoretic because the Hecke indefinite
 modular form at the conformal level $k=1$ is given by the theta
 function $\Theta^1_{1,1}(q) = \eta^2(q)$. More details about the structure
 of the Hecke indefinite modular forms $\Theta^k_{\ell,m}(q)$ associated to
 the affine Kac-Moody algebra of $\rmSU(2)$ can be found in \cite{kp84}.
 This modular form has been discussed previously in a string theoretic
 context in the analysis of a mirror pair involving a rigid
 Calabi-Yau manifold \cite{kls08}.

\underline{$X_3^8(\psi_1,\psi_2) \subset \mathP_{(4,1,1,1,1)}$}

 We consider the singular fiber with $(\psi_1,\psi_2)=(1,1)$
 of the octic family $X_3^8(\psi_1,\psi_2)$. Using the Gauss
 products $\mathG_p^n(u)$ defined above, the $\Om-$motive for the
 Brieskorn-Pham fiber $X_3^8(0)$ can be
 represented as (\ref{omega-motive-orbit}) for $8|(p-1)$ with
 \beq
 M_{\Om}(X_3^8(0)):~~~ (1,1,1,1,1)~ \oplus ~(1,3,3,3,3) ~\oplus ~ (1,5,5,5,5)
            \oplus (1,7,7,7,7).
 \eeq
 This leads to a deformed motive $M_\Om(X_3^8(\psi_1,\psi_2))$ of type
  (\ref{deformed-omega-orbit}) that is generated from the Brieskorn-Pham
  $\Om-$motive by the deformation vectors as
  \beq
  M_\Om(X_3^8(\psi_1,\psi_2)):~~~
       \langle M_{\Om(0,0)}, ~J^1 = (0,4,4,0,0),~
                             J^2 = (0,0,0,4,4) \rangle.
  \eeq
 The resulting motive has rank eight, but degenerates for the $(\psi_1,\psi_2)=(1,1)$ fiber,
 leading to the $L-$function
 \beq
 L_\Om(X_3^{8}(1,1),s)
  = 1 - \frac{22}{5^s} + \frac{18}{13^s} - \frac{94}{17^s}
       + \frac{130}{29^s} - \frac{214}{37^s}
      - \frac{230}{41^s} + \cdots
 \eeq
 with bad prime $p=2$.

This $L-$function agrees with the twist of the $L-$series of a
complex multiplication modular form of weight four and level
$N=32$, given by $f_{4,32} \in S_4(\G_0(32))$ of Table 1. The
twist character is given by $\chi_2$ and the twisted form
$f_{4,32}\otimes \chi_2$ has level $N=64$, leading to
 \beq
 L_\Om(X_3^8(1,1),s)~=~ L(f_{4,32}\otimes \chi_2,s),
 \eeq
 where $\chi_2$ is again a Dirichlet character.

 \vskip .1truein

\underline{More modularity results for CY threefolds}

Other threefolds can be analyzed in a similar way. In Table 3 we
summarize the results for singular fibers in a number of different
weighted one-parameter and two-parameter families of the type
(\ref{special-class}) in weighted ambient spaces
$\mathP_{(w_0,...,w_4)}$.

   \begin{center}
\begin{tabular}{r r }

CY 3fold    &Motivic modular form   \\
\hline

 $X_3^6(1) \subset \mathP_{(2,1,1,1,1)}$
            &$f_{4,108A}(q) \otimes \chi_3 \in S_4(\G_0(432))$  \tabroom \\

 $X_3^6(1,1) \subset \mathP_{(2,1,1,1,1)}$
            &$f_{4,9} \in S_4(\G_0(9))$ \tabroom \\

 $X_3^8(1,1) \subset \mathP_{(4,1,1,1,1)}$
              &$f_{4,32}(q) \otimes \chi_2 \in S_4(\G_0(64))$ \tabroom \\

 $X_3^{12\rmA}(1) \subset \mathP_{(6,2,2,1,1)}$
             &$f_{4,108\rmA}(q) \in S_4(\G_0(108))$ \tabroom \\

 $X_3^{12\rmA}(1,1) \subset \mathP_{(6,2,2,1,1)}$
             &$f_{4,9}(q) \otimes \chi_3 \in S_4(\G_0(144))$  \tabroom \\

 $X_3^{12\rmB}(1) \subset \mathP_{(4,4,2,1,1)}$
             &$f_{4,108\rmB}(q) \otimes \chi_3 \in S_4(\G_0(432))$ \tabroom \\
 \hline
 \end{tabular}
 \end{center}

\centerline{{\bf Table 3.}~{\it Modular fibers in one- and
two-parameter families of Calabi-Yau hypersurfaces.}}

\underline{Remarks} \hfill \break
 1) The motives of the two-parameter threefold families of Table 3 are
 generically of rank eight. The $L-$functions of the fibers at
 $(\psi_1,\psi_2)=(1,1)$ are of CM-type, hence modular by Hecke's
 theory (a list of weight four CM modular forms was constructed in ref.
 \cite{ms06}).

2) The motive of the generic fiber of the one-parameter threefold
family of degree twelve given by $X_3^{12\rmB}(\psi) \subset
\mathP_{(4,4,2,1,1)}$ has rank four. At the singular point
$\psi=1$ this motive degenerates and its $L-$series turns out to
be a twist of the $L-$series of the rank two $\Om-$motive
$M_{\Om(0)}$ of the weighted Fermat hypersurface $X_3^6(0) \subset
\mathP_{(2,1,1,1,1)}$.
 As mentioned above, the $L-$series $L_\Om(X_3^6(0),s)$ is known to be
 modular \cite{rs08}, leading to a form of weight four and
 level $N=108$. The character that produces the correct signs is given
 by $\chi_3$. This leads to the relation
 \beq
 L_\Om(X_3^{12\rmB}(1),s) ~=~ L_\Om(X_3^6(0),s) \otimes \chi_3,
 \eeq
 which via the inverse Mellin transform
  leads to a modular form of the same weight and level $N=432$.

3) The last three families in Table 3 are K3 fibrations whose
generic fibers are given by degree six K3 hypersurfaces
 in the ambient space $\mathP_{(3,1,1,1)}$ for the first two examples, and
 by the degree six K3 hypersurface embedded in
 $\mathP_{(2,2,1,1)}$ for the final entry of the table.
 The Brieskorn-Pham surfaces in these ambient spaces are extremal
 K3 surfaces, i.e. their Picard numbers are maximal. Extremal K3
 surfaces are known to be modular over some number field by the
 work of Shioda and Inose \cite{si77}, but string theoretically only modular
 forms with integral coefficients have been relevant so far.
 This motivates the question about the precise automorphy properties of
 the K3-submotives for the singular threefold fibers.

 To apply our construction to the K3-submotives of the general fiber of
 the family of threefolds, consider the deformation of
 the $\Om-$motive of the Brieskorn-Pham K3 fiber. In the first
 one-parameter family $X_3^{12\rmA}(\psi)$ this leads to the motive
 \beq
  M_{\rmK 3}(X_3^{12\rmA}(\psi)):~~~(1,1,1,2,0)  \oplus (1,1,1,2,0)
         \oplus (1,5,5,10,0) \oplus (1,5,5,4,6),
 \eeq
 which has the $L-$function
 \beq
   L(M_{\rmK 3}(X_3^{12\rmA}(1)),s) ~=~
   L_\Om(X_2^{6\rmA}(0),s),
 \eeq
 and is therefore modular, as noted in our K3 discussion above.

The K3 submotive of the second 1-parameter family
 of K3 fibrations $X_3^{12\rmB}(\psi)$ is given by
  \beq
   M_{\rmK 3}(X_3^{12\rmB}(\psi)):~~~
      (1,1,1,2,0) \oplus (1,1,1,8,6) \oplus (2,2,5,10,0) \oplus (2,2,5,4,6).
 \eeq
  The $L-$function of this motive is again modular, taking the form
  \beq
  L(M_{\rmK 3}(X_3^{12\rmB}(1),s) ~=~
 L_\Om(X_2^{6\rmB}(0),s) ~=~ L(\eta^3(q^2) \eta^3(q^6) \otimes
 \chi_3,s).
 \eeq

\section{CFT aspects of modular singular fibers}

The original motivation for the program of identifying geometric
(i.e. motivic) modular forms with forms arising from the
worldsheet was the idea of using these techniques to provide a
string theoretic construction of spacetime \cite{su02,rs08}. In
this section we consider a second application that arises in the
context of phase transitions in string theory. Such transitions
were originally considered in string theory  via the  splitting
and contraction procedure of ref. \cite{cdls88}. They provide a
connection between Calabi-Yau families with different Hodge
numbers via singular spaces whose degenerations involve singular
points that are nodes. It was shown in \cite{gh88} that the
splitting-contraction construction connects the class of all
complete intersection Calabi-Yau manifolds \cite{cdls88}. The
 geometry of such conifold type spaces was investigated in detail
  in \cite{cgh89, cd90}, and the physical implications were discussed in
\cite{s95,gms95}. Conifold configurations are not the only type of
singularity encountered in the moduli space of Calabi-Yau
varieties, and these other types of degenerations have so far
received little attention.

Related to the question of a better understanding of the physical
nature of more general singularities is the old problem of dealing
with deformations of conformal field theories. Starting from a
rational point in a CFT moduli space, it has proven difficult to
understand the behavior of the deformed theories when one moves
away from the rational point along a marginal direction. It is
therefore of interest to ask whether the modular forms of deformed
Calabi-Yau manifolds with non-nodal singularities can be related
to physical quantities. In this context an observation made above
becomes relevant.

In Section 4 we noted that some motivic modular forms identified
for singular fibers of deformed weighted Fermat hypersurfaces are
identical to modular forms derived from Brieskorn-Pham varieties,
i.e. for smooth fibers at $\psi_J=0$. This immediately implies
that the motivic geometry of such fibers is determined by a
rational conformal field theory, even though it is obtained from a
deformed theory and therefore might have been expected to be a
complicated object in an irrational theory that is difficult to
identify. In our first example, the K3 surface family
$X_2^{6\rmB}(\psi) \subset \mathP_{(1,1,2,2)}$, the motivic
$L-$series at $\psi=1$ was shown to be identical to the
$\Om-$motivic $L-$series of the Brieskorn-Pham fiber
$X_2^{6\rmA}(0) \subset \mathP_{(1,1,1,3)}$. The modular form
$f_\Om(X_2^{6\rmB}(0),q) \in S_3(\G_1(27))$ was determined in
\cite{rs06} as
$f_\Om(X_2^{6\rmA}(0),q)=\eta^2(q^3)\eta^2(q^9)\vartheta(q^3)$.
 This result is surprising for several reasons. First, as noted above, the
 $L-$series of this form can also be viewed as a Hecke $L-$series
  $L_\Om(X_2^{6\rmA}(0),s) = L(\Psi_H^2,s)$ for the Hecke
  character  $\Psi_H = \Psi_{6\rmA}$ defined previously.
  This shows that this $L-$series is the symmetric square of
  the $L-$series defined by the Mellin transform of a weight two modular
  form $f_{2,27}(q) \in S_2(\G_0(27))$. This modular form of weight two turns out to be the
  factor of the weight three form determined by the Dedekind eta function
  \beq
  f_{2,27}(q) = \eta^2(q^3) \eta^2(q^9).
  \eeq
  The origin of the form $f_{2,27}(q)$ is natural from two quite different viewpoints. First, it is
  precisely the modular form of the cubic Fermat elliptic curve
  $E^3$ of eq. (\ref{cubic-fermat-curve})  which appears as a fiber of the elliptic
  family $E^3(\psi)$ defined by the elliptically fibered K3 family $X_2^{6\rmB}(\psi)$.
  From this geometric perspective, the appearance of the string theoretic elliptic modular form
  $f_{2,27}(q)$ is expected.
 Second, the Brieskorn-Pham fiber in the family $X_2^{6\rmB}(\psi)$ has as its
 corresponding Gepner model the GSO projected tensor product $(k=1)^2 \otimes (k=4)^2$ of
 minimal $N=2$ supersymmetric conformal field theories, where $k$ denotes the affine level
 of the model with central charge $c=3k/(k+2)$. Each factor leads to Hecke indefinite
 modular forms
  \beq
 \Theta^k_{\ell,m}(\tau)
  = \sum_{\stackrel{\stackrel{-|x|<y\leq |x|}{(x,y)~{\rm
or}~(\frac{1}{2}-x,\frac{1}{2}+y)}}{\in
\ZZ^2+\left(\frac{\ell+1}{2(k+2)},\frac{m}{2k}\right)}} \rmsign(x)
e^{2\pi i \tau((k+2)x^2-ky^2)},
 \eeq
 which are related to the parafermionic Kac-Peterson string functions $c^k_{\ell,m}(\tau)$ as
 \beq
  \Theta^k_{\ell,m}(\tau) ~=~ \eta^3(\tau) c^k_{\ell,m}(\tau).
 \eeq
 These string functions, in combination with standard theta functions, completely determine
  the characters of the $N=2$ superconformal model \cite{g88}. For affine level $k=1$, there exists
  a unique Hecke indefinite modular form, given by
  \beq
  \Theta^1_{1,1}(q) = \eta^2(q).
  \eeq
  This then leads to the worldsheet interpretation of the motive of the
  fiber $X_2^{6\rmB}(1)$ by recognizing that the weight two building block
  can be written as
  $f_{2,27}(q) ~=~ \Theta^1_{1,1}(q^3) \Theta^1_{1,1}(q^9)$.

 Other examples can be analyzed in a similar way, and in Table 4 we summarize
 those results that explicate the emergence of modular forms that arise from
 deformed $\Om-$motives in singular configurations, but which also appear
 as motivic forms in Calabi-Yau varieties of weighted Fermat type.
 The motivic $L-$functions of all the fibers $X(\psi)$ in Table 4 are either identical to
 motivic $L-$functions of Brieskorn-Pham hypersurfaces $X'(0)$ or twists of such $L-$series
 that we have previously shown to be modular in terms of string theoretic
 modular forms \cite{rs06,rs08}. Thus we have
 \beq
 L_\Om(X(\psi),s) ~=~  L(f_\Om(X'(0))\otimes \chi,s),
 \eeq
 where $\chi$ is a (possibly trivial) Dirichlet character, and $f_\Om(X'(0)),q)$
 is a known modular form $f_\Om(X'(0),q) \in S_w(\G_0(N))$, where $w=n+1$
 and $N$ is the level.
 These modular forms are all of CM-type, hence modularity follows directly from
 Hecke's theory. The string theoretic
 interpretation of these forms proceeds again in terms of the Hecke indefinite
 modular forms defined above.
  \begin{center}
\begin{tabular}{r r r}

 CY fiber $X_n^d(1)$       &BP variety $X_n^d(0)$
                                        &Modular form   \tabroom \\
\hline

 $X_2^{6\rmB}(1) \subset \mathP_{(2,2,1,1)}$
              &$X_2^{6\rmA}(0) \subset \mathP_{(3,1,1,1)}$
              &$f_{3,27}%  \in S_3(\G_1(27))
                $  \tabroom \\

 $X_2^{8}(1) \subset \mathP_{(4,2,1,1)}$
             &$X_2^{4}(0) \subset \mathP_3$
             &$f_{3,16} \otimes \chi_2 % \in S_3(\G_1(64))
               $ \tabroom \\

 $X_2^{12\rmA}(1) \subset \mathP_{(6,4,1,1)}$
             &$X_2^{6\rmA}(0) \subset \mathP_{(3,1,1,1)}$
             &$f_{3,27}(q)\otimes \chi_3$ \tabroom \\

 \hline

 $X_3^{12\rmB}(1) \subset \mathP_{(4,4,2,1,1)}$
            &$X_3^{6}(0) \subset \mathP_{(2,1,1,1,1)}$
             &$f_{4,108\rmB}(q)\otimes \chi_3$ \tabroom \\

        \hline
 \end{tabular}
 \end{center}

\centerline{{\bf Table 4.}~{\it Relations between modular forms of singular fibers
and forms of Brieskorn-Pham fibers.}}

 The results described for the examples above, and their extension to the varieties
 in Table 4, show that the structure of these singular fibers is determined by
  the modular forms that emerge at the smooth Brieskorn-Pham point of the family
  corresponding to the Gepner point. The existence of such $L-$correlated pairs
  suggests that the singular phases that occur in the families $X_n^d(\psi_1,\psi_2)$
  considered here admit a physical
  interpretation and that the conformal field theory is much
  better under control than previously thought.

\section{Motivic dimensional transmutation}

In the previous section we have shown that singular Calabi-Yau
varieties with high Milnor numbers may lead to geometric modular
forms that can be identified with string theoretic forms on the
worldsheet. In the examples discussed so far, the degeneration of
the motive only affects the rank of the motive, not its weight.
The weight of the motive of a smooth variety is a characteristic
that is determined by the degree of the cohomology associated with
the motive. In the case of the $\Om-$motive of smooth Calabi-Yau
varieties this is the intermediate cohomology, in which case the
weight $\rmwt(M_\Om)$ is given by $\rmwt(M_\Om)= n$, where $n$ is
the complex dimension of the variety. For the special case of
smooth Calabi-Yau varieties the weight of the $\Om-$motive can
therefore be viewed as a proxy for the dimension of the space. The
weight of the resulting modular form $w(f)$ is then given by
 $w(f)=\rmwt(M_\Om)+1$.
 In this  section we will discuss examples in which the degeneration
 of the motive induced by the singularities reduces not only the rank but also
 the weight, leading to a kind of dimensional
 transmutation of the motive at the singular fibers of high Milnor number.

Perhaps the simplest example in which the phenomenon of motivic
dimensional transmutation happens is the quartic K3 surface of
type (\ref{2-parameter-family-class}) with $\psi_1=0$,
$\psi_2=\psi$, and $b=2$. In this case the deformed motive
 \beq
 M_{\Om}(X_2^4(\psi)):~~~~(1,1,1,1)\oplus (1,1,3,3) \oplus (3,3,3,3) \oplus (3,3,1,1)
 \eeq
 for the generic fiber is of rank four, the same rank as for the motives of
 the generic fibers of all the
 one-parameter K3 families considered in the previous section. The $L-$function of this
 motive is given by
  \beq
  L_\Om(X_2^4(1),s) ~=~ 1 + \frac{2}{5^s} - \frac{6}{13^s} +
  \frac{2}{17^s} + \frac{10}{29^s} + \frac{2}{37^s} + \cdots .
  \eeq
 This shows that this $\Om-$motivic $L-$function of a complex surface is in fact identical
 to the $L-$series of the quartic weighted Fermat elliptic curve
 $E^4 \subset \mathP_{(2,1,1)}$
 \beq
 L_\Om(X_2^4(1),s) ~=~ L(E^4,s).
 \lleq{quartic-curve}
 The modular form of the curve $E^4$ was determined in \cite{ls04} to
 be given by the weight two form at level 64
 \beq
 L(E^4,s) ~=~ L(\eta^2(q^4)\eta^2(q^8) \otimes \chi_2, s),
 \eeq
 where the expansion of $f_{2,32}(q) = \eta^2(q^4)\eta^2(q^8) \in S_2(\G_0(32))$
 is given in Table 1. It follows that this $L-$series is string theoretic at conformal
 level $k=2$, since the modular forms $f_{2,32}$ can be written in terms of the Hecke
 indefinite modular forms as
 $\Theta^2_{1,1}(q) = \eta^2(q)\eta^2(q^2)$. This leads to the string theoretic identification of
 the motivic modular form (\ref{quartic-curve}), and therefore of the highly degenerate motive
 coming from the singular fiber of the quartic K3 family $X_2^4(\psi)$.

For Calabi-Yau threefolds similar phenomena emerge. Consider, for
example, the two-parameter family of threefolds
 $X_3^{8\rmB}(\psi_1,\psi_2) \subset \mathP_{(2,2,2,1,1)}$ of type
 (\ref{2-parameter-family-class}) with $a=2$ and $b=4$. The deformed
 $\Om-$motive for the general fiber is of rank eight
 \beq
 M_{\Om}(X(\psi_1,\psi_2)):~~~~
  \langle M_{\Om(0)}, ~J^1 = (0,2,2,0,0),~J^2=(0,0,0,4,4)\rangle,
 \eeq
 the same rank as for all deformed motives of the generic fibers of the two-parameter
 families discussed in the previous section.
 % (1,1,1,1,1) \oplus (1,4,4,1,1) \oplus (1,1,1,7,7) \oplus (1,4,4,7,7)
 % (1,5,5,5,5) \oplus (1,2,2,5,5) \oplus (1,5,5,11,11) \oplus (1,2,2,11,11)
 The $L-$function of this motive is identical to the motivic $L-$function of the
 octic K3 surface $X_2^{8}(1) \subset \mathP_{(4,2,1,1)}$
  \beq
  L_\Om(X_3^{8\rmB}(1,1),s) ~=~ L_\Om(X_2^8(1),s),
  \eeq
 considered earlier in this paper.

 We have seen in Section 5 that the $L-$series of $X_2^8(1)$ is modular, given in fact
 by the $\chi_2-$twist of the $L-$function of the Fermat K3 surface $X_2^{4}(0)$. The resulting
 CM modular form is therefore string modular according to the results in \cite{rs06}.
 A second Calabi-Yau threefold fiber of this type appears in the twoparameter family
 $X_3^{12\rmC}(\psi_1,\psi_2) \subset \mathP_{(4,3,3,1,1)}$,
 where we set $(\psi_1,\psi_2)=(1,1)$.
 Again, the $L-$series is identical to that of a K3 surface, in this
 case $X_2^{12}(1) \subset \mathP_{(6,4,1,1)}$, which in turn is the twist of the
 $L-$series of the degree six weighted Fermat K3 hypersurface
 $X_2^{6\rmA}(0)\subset \mathP_{(3,1,1,1)}$. The $\Om-$motive of this latter surface is
 known \cite{rs06} to be modular with $L_\Om(X_2^{6\rmA}(0),s) ~=~
 L(f_{3,27},s)$.  We collect the modular data of these examples in Table 5.
 \begin{center}
 \begin{tabular}{l c | l  c}
 Fiber         &Modular form
               &BP variety     &Modular form \\
 \hline

 $X_2^4(1)\subset \mathP_3$    &$f_{2,32} \otimes \chi_2$
                        &$E^{4} \subset \mathP_{(2,1,1)}$
                        &$f_{2,32}\otimes \chi_2$   \tabroom \\

 $X_3^{8\rmA}(1,1) \subset \mathP_{(2,2,2,1,1)}$
          & $f_{3,16}\otimes \chi_2$
                                 &$X_2^{4}(0) \subset \mathP_{3}$
                    &$f_{3,16}$   \tabroom \\

 $X_3^{12\rmC}(1,1) \subset \mathP_{(4,3,3,1,1)}$
                               &$f_{3,27}\otimes \chi_3$
                                &$X_2^{6\rmA}(0) \subset \mathP_{(3,1,1,1)}$
                    &$f_{3,27}$    \tabroom \\
 \hline
 \end{tabular}
 \end{center}

 \centerline{{\bf Table. 5.}~{\it Examples with modular weight reduction, leading to
  forms $f_{w,N}$ with $w<n+1$.}}

The results discussed here and in the previous section show that
both the effective rank and the modular weight of a motive can
change dramatically as the family varies over the moduli space. In
the previous section we saw examples in which the effective rank
of the motive changed such that a motive that was expected to be
automorphic of high degree reduced to something modular with
weight precisely matching that expected from the dimension of the
variety (or rather the level of the intermediate cohomology). In
contrast, in this section we encountered examples in which not
only was the effective rank reduced, but also the modular weight.

 The results discussed here not only show that a family of high
 rank motives can become modular at singular points in the moduli space,
 but that the resulting modular forms can arise from lower-dimensional
 varieties. These modular forms retain their string theoretic nature
 even in such highly degenerate situations.
 The phenomenon of dimensional transmutation of motives of singular
 Calabi-Yau spaces thus suggests that such singular points in the
 moduli space should define transitions between string vacua of
 different dimensions. It might therefore be possible to embed
 into string theory considerations of earlier papers that concerned different
decompactification mechanisms in lower-dimensional ad-hoc models
\cite{gm04, cjr09, bsv09}.

\section{Concluding remarks}

In this paper we have taken the first steps in generalizing the
higher-dimensional string modularity results obtained in
\cite{rs06,rs08,kls08} for Brieskorn-Pham type varieties to families
of Calabi-Yau spaces. This establishes that the program of
constructing an emergent
 spacetime within string theory via modular methods transcends the
 framework of exactly solvable rational conformal field theories
 and generalizes to deformations of such theories via marginal operators.

An unexpected result which indicates the usefulness of the modular
techniques is that the string theoretic interpretation shows that
completely different fibers in the moduli space of K3 surfaces
lead to identical modular forms. Since these modular forms
determine the structure of the underlying conformal field theory,
this shows that the worldsheet theories lead to identical modular
structures, although the geometry is rather different. This
implies that deformations of rational conformal field theories
inherit at least part of the structure of the rational points in
moduli space, in particular as far as the structure of spacetime
is concerned.

Furthermore, rather surprisingly the modular forms associated to
the deformed $\Om-$motives of singular fibers can be identical to
modular forms arising from the $\Om-$motives of smooth
Brieskorn-Pham manifolds. The construction of these forms in terms
of the forms of the worldsheet suggests that string theory might
be viable on spaces with singularities of high multiplicity and
that such configurations may be transit points of phase
transition, similar to conifolds in the case of node-type
singularities.

Finally, we have observed the phenomenon that at certain singular
fibers with higher multiplicity singular points the $L-$function
can degenerate so that the motive reduces not only its rank but
also its weight, leading to what might be called motivic
dimensional transmutation. Even in these cases, the modular forms
that appear in our examples admit a string theoretic
interpretation.

It would be of interest to see whether the conformal field
theoretic techniques developed by Wendland \cite{kw04} over the
past years can be used to illuminate the CFT structure of the
singular fibers described in this paper.

\vskip .1truein

{\bf ACKNOWLEDGEMENT.}

M.L. and R.S. are supported in part by the National Science
Foundation under Grant No. 0969725. The work of R.S. was also
funded in part by IUSB Faculty Research Grants. The authors are
grateful to the Oberwolfach Mathematical Institute for support
through a RIP grant in 2008. R.S. also thanks Klaus Hulek at the
Institute for Algebraic Geometry at the Universit\"at Hannover and
Katrin Wendland at the Universit\"at Augsburg for hospitality. It
is a pleasure to thank Michael Flohr, Klaus Hulek, Emanuel
Scheidegger and Katrin Wendland for discussions.

\vskip .5truein

\baselineskip=17pt
\parskip=0.00truein

%\begin{small}

%\end{small}

\end{document}